\documentclass[aps,prl,reprint,groupedaddress,amsfonts,amsmath,showpacs]{revtex4-1}

\usepackage{xcolor}
\definecolor{darkgreen}{rgb}{0,0.6,0}
\definecolor{cyan}{rgb}{0,0.7,0.8}
\usepackage[colorlinks=true,citecolor=darkgreen,urlcolor=cyan]{hyperref}
\usepackage{graphicx}

\newcommand{\mrm}[1]{\mathrm{#1}}

\newcommand{\mbb}[1]{\mathbb{#1}}

\newcommand{\Eref}[1]{Eq.~(\ref{#1})}

\newcommand{\fref}[1]{Fig.~\ref{#1}}

\newcommand{\pfref}[1]{\protect{Fig.~\ref{#1}}}

\newcommand{\rmi}{\mathrm{i}}
\newcommand{\ra}{\rangle}
\newcommand{\la}{\langle}

\newcommand{\rcite}[1]{Ref.~\onlinecite{#1}}
\newcommand{\rcites}[1]{Refs.~\onlinecite{#1}}
\newcommand{\prcite}[1]{Ref.~\protect{\onlinecite{#1}}}

\newcommand{\bonderson}{Refs.~\onlinecite{bonderson2007} and \onlinecite{bonderson2008}}
\newcommand{\ff}{\rho}

\begin{document}

\title{Phase diagram for hard-core $\mbb{Z}_3$ anyons on the ladder}

\author{Robert N. C. Pfeifer}
\email[]{robert.pfeifer@mq.edu.au}
\affiliation{Dept. of Physics \& Astronomy, Macquarie University, Sydney, NSW 2109, Australia}

\date{\today}

\begin{abstract}
Studies of free particles in low-dimensional quantum systems such as two-leg ladders provide insight into the influence of statistics on collective behaviour. The behaviours of bosons and fermions are well understood, but two-dimensional systems also admit excitations with alternative statistics known as anyons. Numerical analysis of hard-core $\mathbb{Z}_3$ anyons on the ladder reveals qualitatively distinct behaviour, including a novel phase transition associated with crystallisation of hole degrees of freedom into a periodic foam. Qualitative predictions are extrapolated for all Abelian $\mathbb{Z}_q$ anyon models.
\end{abstract}


\maketitle

The study of low-dimensional quantum systems through both theory and experiment provides a unique window into the collective behaviours of quantum many-body systems \cite{giamarchi2004,bloch2008,cazalilla2011}. %
Competition between interaction terms, hopping terms, filling fractions, and geometrical frustration may give rise to an immensely broad range of different phases of matter, and transitions may be observed between regimes which are insulating, conductive, or superconductive, ordered or disordered, and many more. Moreover, %
the statistics of the particles in a model may significantly affect its phase diagram \cite[e.g.][]{girardeau1960,giamarchi1988,fisher1989,albus2003,lewenstein2004,crepin2011}. %

Other than the 1D chain, one of the simplest low-dimensional quantum systems is that of a hard-core non-interacting gas on a two-leg ladder. An infinitely long ladder may be considered to be a quasi-1D system, although at incomplete fillings it is possible for particles to braid around one another and thus the phase diagram is affected by particle statistics. %
The behaviours of both fermions and hard-core bosons on the two-leg ladder have been extensively studied, and are known to generate rich phase diagrams dependent both on the ratio of the rung and leg hopping terms and on the value of the chemical potential \cite{crepin2011}. However, these two examples are not exhaustive as there are other choices for particle statistics in one- and two-dimensional systems. These \emph{anyonic} statistics are of fundamental interest as exotic and as-yet poorly-understood phases of matter. Efforts to understand their behaviour are also increasingly driven by 
their roles in phenomena such as 
the fractional quantum Hall effect (FQHE), \cite{laughlin1983,halperin1984,fradkin1989,read1999,xia2004,nayak2008,pan2008,stern2010,sanghun-an2011}, 
and %
as exotic excitations in superconducting heterostructures and nanowires \cite{read2000,kitaev2001,alicea2012,beenakker2013,clarke2013,stanescu2013,nadj-perge2014,mong2014,klinovaja2014,vaezi2014}.

In this paper I consider a simple %
example of anyonic statistics, the {$\mbb{Z}_3$ anyon} model, which is realised in the FQHE at $\nu=\frac{2}{3}$ %
\cite{halperin1984} and in coupled wire constructions %
\cite{meng2015,gorohovsky2015}.
Whereas a bosonic system is unchanged under pair exchange and a fermionic system picks up a factor of $-1$ when exchanging particles which are nearest neighbours with respect to some linearly-ordered basis, a system of anyons may %
undergo a unitary transformation.
For $\mbb{Z}_3$ anyons, this transformation is a phase and is a third root of unity. 

The study of anyonic systems is computationally challenging, as their non-trivial exchange statistics give rise to a form of the sign problem \cite{loh1990,troyer2005}. 
However, the class of simulations known as tensor network algorithms are known to be relatively resilient against this effect \cite{corboz2009,corboz2010,corboz2010a,corboz2010b,bauer2011}. Recent work on the incorporation of both Abelian and non-Abelian symmetries into tensor network algorithms \cite{mcculloch2000,mcculloch2002,singh2010,singh2011,singh2012} admits a generalisation to arbitrary anyonic systems \cite{pfeifer2010,pfeifer2011a,pfeifer2012a}, and this has led to the recent development of anyonic tensor network algorithms such as anyonic Time-Evolving Block Decimation (TEBD) \cite{singh2014} and anyonic Density Matrix Renormalisation Group (DMRG) \cite{pfeifer2015,pfeifer2015a}. In the present paper, the anyonic version of the infinite DMRG algorithm \cite{white1992,white1993,schollwock2011,pfeifer2015a} is used to study $\mbb{Z}_3$ anyons on the two-leg infinite ladder. Performance is enhanced using %
precomputation techniques 
\cite{singh2011}.

The $\mbb{Z}_3$ anyon model has two particle species, also known as charges, which are customarily denoted~1 and~2, and a vacuum sector, denoted~0. Fusion of particles is equivalent to summation of charges, represented by addition modulo~3, $a\times b\rightarrow (a+b)~\mrm{mod}~3$, and exchange of two neighbouring charges $a$ and $b$ in a linearly-ordered basis attracts a factor of $R^{ab}$ or $(R^{ab})^*$ where $R^{ab}=\mrm{exp}({\frac{2\pi\rmi ab}{3}})$. Whether this factor is conjugated depends on which particle passes behind, and which in front of, the other, with respect to some oriented, linearly-ordered basis. It follows from these definitions that species~2 is the antiparticle of species~1, with the model being invariant under simultaneous interchange of charge labels~1 and~2 and time reversal (%
complex conjugation). In this paper we consider only a single species of hard-core $\mbb{Z}_3$ particle on the ladder \footnote{A study of \protect{$\mbb{Z}_q$} anyons on the chain may be found in \prcite{tang2015}.}, with the resulting phase diagram being independent of our choice. For definiteness we may consider this to be species~1.

As with fermions we begin by imposing a linear ordering on the sites of the 1D ladder:
\begin{equation}
\raisebox{-15pt}{\includegraphics[width=2in]{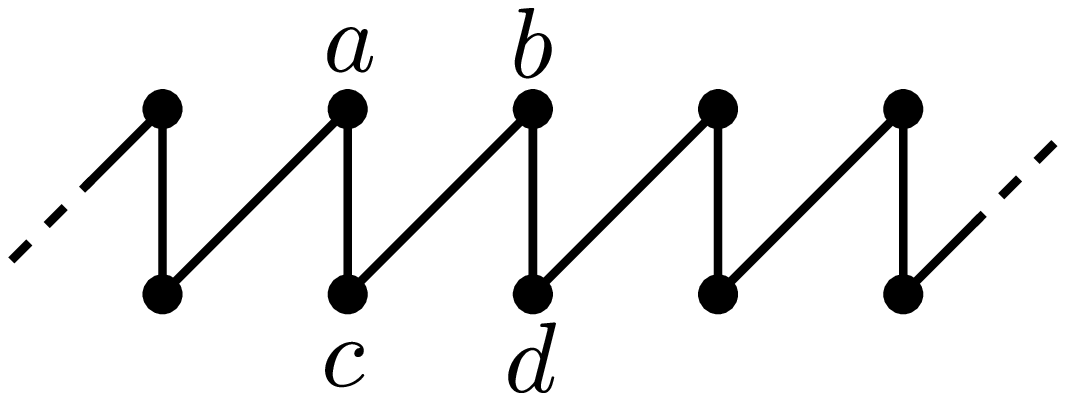}} %
\label{eq:ordering}
\end{equation}
This order is supplemented by an orientation---we specify that this basis is viewed from the lower side of the diagram. Thus if a particle hops from site $a$ to site $b$ in the above diagram, it passes behind site $c$, and if one hops from $c$ to $d$, it passes in front of site $b$. If a particle is also present on the site being hopped past, then this attracts a phase factor $R^{11}$ or $(R^{11})^*$.
Using a graphical notation similar to that in \bonderson{} and the ordering of \Eref{eq:ordering}, the Hamiltonian for hard-core particles with nearest neighbour hopping and a chemical potential may be written
\begin{equation}
\begin{split}
\hat H=&-t\!\!\!\!\!\!\!\sum_{i\in\mrm{upper~leg}}\left(~\raisebox{-12.9pt}{\includegraphics[width=1.2cm]{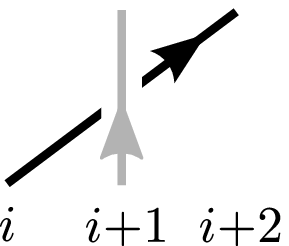}}+\raisebox{-12.9pt}{\includegraphics[width=1.2cm]{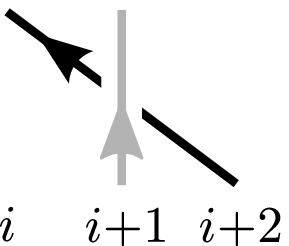}}~\right)\\
&-t\!\!\!\!\!\!\sum_{i\in\mrm{lower~leg}}\left(~\raisebox{-12.9pt}{\includegraphics[width=1.2cm]{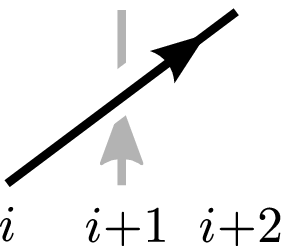}}+\raisebox{-12.9pt}{\includegraphics[width=1.2cm]{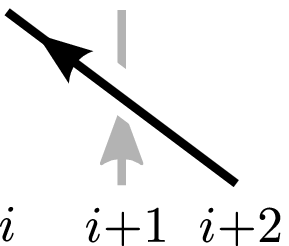}}~\right)\\&-t_\perp\sum_i\left(\,\raisebox{-14.5pt}{\includegraphics[width=0.8cm]{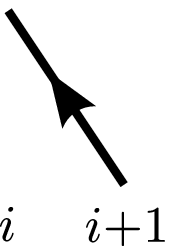}}+\raisebox{-14.5pt}{\includegraphics[width=0.8cm]{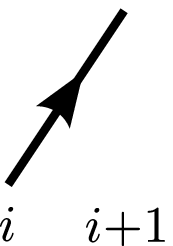}}\,\right)-\mu\sum_i\!\!\!\raisebox{-14.5pt}{\includegraphics[width=0.8cm]{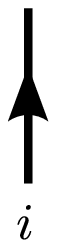}}
\end{split}\label{eq:H}
\end{equation}
where a solid line indicates the presence of a particle (charge~1), the absence of a line indicates the vacuum sector (charge~0), and a grey line may be either.

\begin{figure}
\includegraphics[width=\columnwidth]{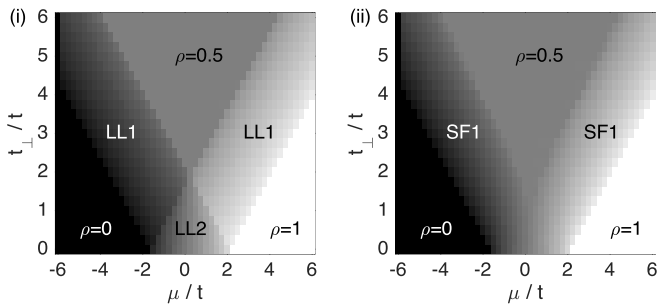}
\caption{(i)~Phase diagram for fermions on the 2-leg ladder. (ii)~Phase diagram for hard-core bosons on the 2-leg ladder. Greyscale denotes filling fraction, from~0 (black) to~1 (white). Labels LL1 and LL2 indicate 1-band and 2-band Luttinger liquids respectively in the fermionic system, SF denotes a superfluid phase for the hard-core bosons, and $\ff{}=\ldots$ signifies an insulating phase with the specified filling fraction. The $\ff{}=\frac{1}{2}$ phase is a band insulator for fermions, and a rung-Mott insulator for hard-core bosons. These results reproduce phase diagrams previously presented in \prcite{crepin2011}.\label{fig:results1}}
\end{figure}
\begin{figure}
\includegraphics[width=\columnwidth]{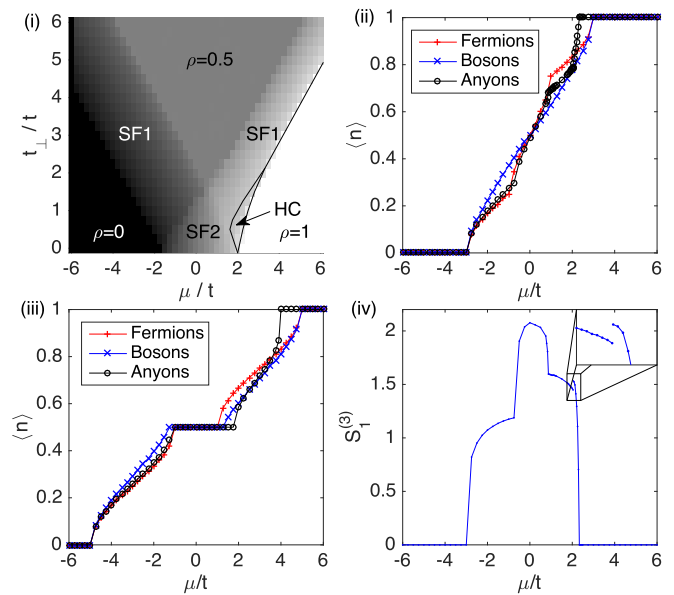}
\caption{(Colour online) (i)~Phase diagram for hard-core $\mbb{Z}_3$ anyons on the 2-leg ladder. Labels SF1 and SF2 indicate superfluid regimes analogous to fermionic 1-band and 2-band Luttinger liquids, and $\ff{}=\ldots$ signifies an insulating phase with the specified filling fraction. HC labels the novel ``hole crystal'' phase described in the text, with thin black lines indicating the approximate boundaries of %
this phase. %
(ii)~Section through diagram~(i) at $t_\perp/t=1$, showing $\la n\ra$ as a function of $\mu/t$. Results for $\mbb{Z}_3$ anyons are indicated by $\circ$, with results for fermions and hard-core bosons being extracted from \pfref{fig:results1} and denoted $+$ and $\times$ respectively. (iii)~Section through diagram~(i) at $t_\perp/t=3$%
. Key as for diagram~(ii). (iv)~Three-rung von Neumann entanglement entropy $S_1^{(3)}$ as a function of $\mu/t$ for $t_\perp/t=1$.
A discontinuity in $S_1^{(3)}$ at $\mu/t\approx 2.04$ shows the SF1/HC phase transition to be first-order. %
\label{fig:results2}}
\end{figure}

%
In simulating this system the accessory site variation of the iDMRG algorithm discussed in \rcite{pfeifer2015a} provides exceptionally fast convergence towards the ground state, and comparison with \rcite{crepin2011} shows that for a refinement parameter $D=200$, an accurate picture of the phase diagram for fermions and hard-core bosons emerges after as few as~25 iterations, as shown in \fref{fig:results1}, with the insulating and the Luttinger liquid or superfluid phases of these models being clearly visible.
Application of the same technique to a system of $\mbb{Z}_3$ anyons yields the phase diagram shown in \fref{fig:results2}(i). 
Individual sections through this diagram were also computed with up to~300 iterations and $D=400$ and were confirmed to show no qualitative changes, with only minor quantitative changes on the order of one percent filling.
The three-rung von Neumann entanglement entropy $S_1^{(3)}$ [\fref{fig:results2}(iv)] was computed from the $3$-rung reduced density matrix $\hat \rho^{(3)}$ using the anyonic entanglement entropy formalisms of \rcites{pfeifer2014,kato2014}. As $\mbb{Z}_3$ anyons have a quantum dimension of 1, these reduce to the usual expression
\begin{equation}
S_1^{(3)}=-\mrm{Tr}[\hat\rho^{(3)}\log{\hat\rho^{(3)}}].
\end{equation}

The most obvious feature of \fref{fig:results2}(i) is the loss of particle/hole duality, corresponding to reflection symmetry of the phase diagram about $\mu/t=0$. While this property holds for both fermions and hard-core bosons, its absence is unsurprising for $\mbb{Z}_3$ anyons: 
For both fermions and bosons %
this property may be identified with the existence of a $\mbb{Z}_2$ symmetry associated with conservation of parity. Starting at the $\mu/t=0$ half-filling state, any small perturbation introducing or removing a particle from the half-filled state causes interchange of the charge~0 and~1 sectors of the MPS chain, with a corresponding shift in the chemical potential being required to stabilise this change. For $\mbb{Z}_3$ anyons, this symmetry is broken. The dual of charge~1 is now charge~2, and addition or deletion of a particle on a site causes the charge sectors to increment or decrement by one, modulo~3. Thus the defects introduced in the MPS chain by insertion and deletion of an additional particle are no longer equivalent, and symmetry of the phase diagram about $\mu/t=0$ and $\ff{}=\frac{1}{2}$ would be surprising, rather than expected.

Further qualitative insight into the quantitative behaviour revealed in \fref{fig:results2}(i) may be obtained by comparing the filling fraction as a function of $\mu/t$ 
with that for fermions and hard-core bosons, at specific values of $t_\perp/t$. %
Transects performed at $t_\perp/t=1$ and $t_\perp/t=3$, as in \rcite{crepin2011}, %
are enlightening and are shown in \fref{fig:results2}(ii) and~(iii). 

Consider first \fref{fig:results2}(iii), $t_\perp/t=3$, where the behaviour of $\mbb{Z}_3$ anyons at filling fractions $0\leq\ff{}\leq \frac{1}{2}$ is seen to closely resemble that of fermions, though with a slight deviation towards higher filling fractions in the first conducting phase. 
As with fermions the repeated exchange of $\mbb{Z}_3$ anyons leads to destructive interference, although a sum over three terms rather than two is required in order for the contributing diagrams to sum to zero. %
During real-time evolution of a hard-core gas in the dilute limit, particles are relatively free to braid around each other, and two particles in sufficiently close proximity to undergo pair exchange may expect their future light cones to overlap for a significant interval before encountering a third particle. The longer this interval, the greater the relative extent of the overlap: 
\begin{center}%
\includegraphics[width=\columnwidth]{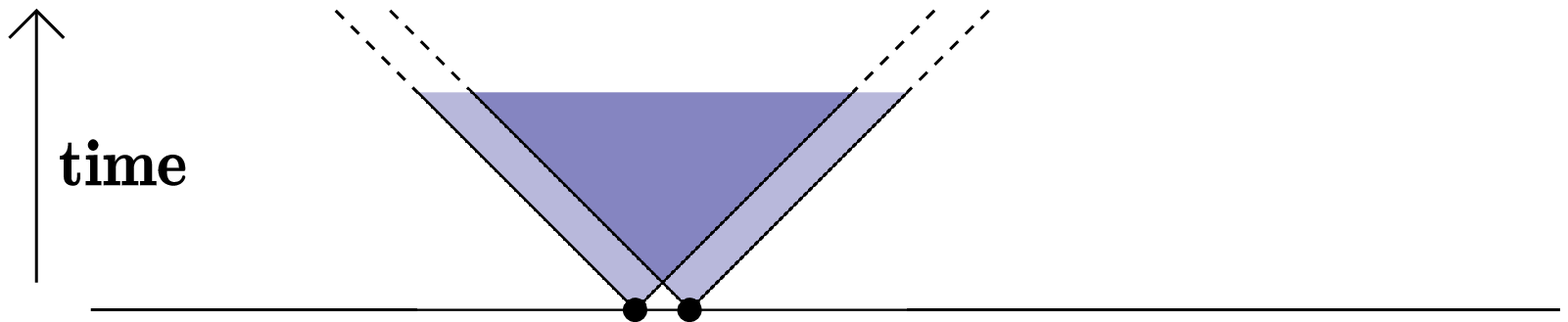}
\end{center}%
For a sufficiently dilute gas the time to encounter another particle approaches infinity, under which circumstances the overlap of the future light cones approaches unity, and the number of times modulo~3 that a given pair of anyons undergoes exchange is equally likely to be~0,~1, or~2. Full destructive interference, leading to precise equivalence of filling fractions for fermions and $\mbb{Z}_3$ anyons, is therefore approached in the limit of a sufficiently dilute anyon gas \footnote{The argument extends to larger isolated groups of particles by summing over arbitrary permutations on the group.}. 
At finite but small filling fractions, larger numbers of exchanges become less likely than smaller numbers for any given pair encounter, so $P(2)<P(1)$, and a small residual term arises from zero- and single exchanges not matched by double exchanges. It is anticipated that for increasing anyon species $q$ in the set of models $\{\mbb{Z}_q\}$ and fixed filling fraction $\ff{}$ this residual term will increase, with the limit $q\rightarrow\infty$ corresponding to hard-core bosons where pair exchanges are not associated with any destructive interference. Thus the $\mbb{Z}_q$ anyons are anticipated to interpolate between fermions and hard-core bosons in the left half of \fref{fig:results2}(iii). 

As with both bosons and fermions, a plateau is observed at $\ff{}=\frac{1}{2}$. Although the behaviour of $\mbb{Z}_3$ anyons still strongly resembles that of fermions in this regime, this plateau must be considered a rung-Mott insulator as the addition of further particles is not inhibited by lack of available states in the band structure, but rather by the combination of infinite on-site repulsive interaction and the energy penalty associated with obstruction of the hopping terms in the Hamiltonian. At the low-$\mu/t$ end of the plateau, the anyons' pseudo-fermionic character predominates. However, at the high-$\mu/t$ end of the $\ff{}=1/2$ plateau the insulating phase is seen to be more stable than its bosonic and fermionic counterparts, as is the plateau at $\ff{}=1$.

In this higher filling regime, $\frac{1}{2}<\ff{}<1$, several effects apply. First, the introduction of additional particles reduces the amount of hopping which can take place, and the increase in chemical potential must be sufficient to overcome the associated energy penalty. However, this effect applies equally to all species and so does not explain %
the unique behaviour of $\mbb{Z}_3$ anyons. 

Beyond $\ff{}=1/2$ it is useful to think of the holes as the free excitations on an otherwise full lattice, with a rung occupancy of $n$ corresponding to $2-n$ holes. For fermions particle/hole duality applies, allowing the holes to be viewed as the carriers of current above $\ff{}=1/2$. Adding particles is therefore favourable as it decreases the amount of destructive interference associated with hole exchange. Hard-core bosons also exhibit particle/hole duality, but removal of holes is neutral with regards to interference so a higher chemical potential is required to leave the $\ff{}=1/2$ plateau. Finally, for $\mbb{Z}_3$ anyons the holes may still be thought of as the free excitations at higher $\ff{}$ but fully-occupied rungs now carry a charge of 2. In this context even relatively simple hole trajectories such as
\begin{center}%
\includegraphics[width=\columnwidth]{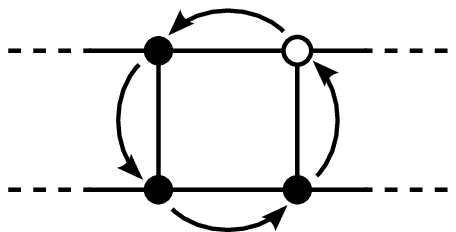}
\end{center}%
can generate destructive interference, opposing transition into a hole-mediated conducting regime. Similarly, starting from the $\ff{}=1$ plateau and decreasing $\mu/t$ the $\ff{}=1$ plateau is likewise seen to be more extensive
as destructive interference %
again opposes the introduction of holes. %
Noting that $\mbb{Z}_2$ is a special case for which the saturation of a rung with particles yields a total rung charge of 0, it is expected that for all \emph{other} $\mbb{Z}_q$, i.e.~$\mbb{Z}_q|_{3\leq q\in\mbb{Z}^+}$, destructive interference will inhibit departure from the $\ff{}=1/2$ and $\ff{}=1$ plateaus. However, this effect will decrease with increasing $q$ as the phase angle $R^{11}%
$ which is associated with braiding decreases, attaining hard-core boson behaviour in the limit $q\rightarrow\infty$.

In \fref{fig:results2}(ii) the behaviour of $\mbb{Z}_3$ anyons is again fermion-like for $\ff{}<1/2$ with deflection towards the hard-core boson line due to a reduction in destructive interference. 
To the extent that a quasi-fermionic description applies, the weakening of destructive interference may be viewed as deforming %
the effective
band structures, %
delaying transition from the one-band superfluid phase (SF1) to the two-band superfluid phase (SF2) and permitting the attainment of %
higher filling fractions in SF1.

For $t_\perp/t=1$ there is no insulating phase at $\ff{}=\frac{1}{2}$ due to overlap in the effective band structure, and the behaviour of the anyonic model continues to 
qualitatively mimic that of the fermionic model right up to $\ff{}\approx 0.78$. The role of destructive interference decreases rapidly, however, as a higher particle density provides an obstruction to repeated braiding, and this causes the $\mbb{Z}_3$ line to shift further away from the fermion line and towards the hard-core boson line. 

\begin{figure}
\includegraphics[width=\columnwidth]{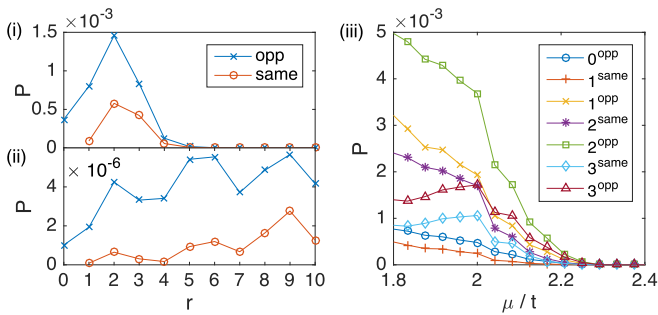}
\caption{(Colour online) Probabilities for different separations of nearest-neighbour holes in the hole-crystal phase with $t_\perp/t=1$, at (i)~$\mu/t=2.06$ and (ii)~$\mu/t=2.28$. Labels ``opp'' and ``same'' indicate whether %
holes are on opposite or the same rungs respectively.
(iii)~Probabilities for different separations of nearest-neighbour holes as a function of $\mu/t$, in superfluid and hole-crystal phases with $t_\perp/t=1$.
\label{fig:results3}}
\end{figure}

At approximately $\mu/t=2.04,~\ff{}=0.78$
there 
is a phase transition from a conventional superfluid phase with predominantly zero or one vacancies per rung to a phase in which the holes are organised in a regular crystalline structure (labelled HC). 
Discontinuity of the 3-site von Neumann entanglement entropy shows that this phase transition
is first-order [\fref{fig:results2}(iv)].
In the HC phase, %
configurations are favoured where sequential pairs of holes are located on opposite legs of the ladder, at separations which admit bosonic exchange statistics. In the vicinity of the critical point the typical separation is 2 rungs: \raisebox{-3.33pt}{\includegraphics[width=0.5cm]{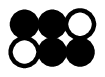}} or \raisebox{-3.33pt}{\includegraphics[width=0.5cm]{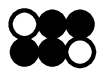}}, with these configurations being collectively denoted $2^\mrm{opp}$ 
($2^\mrm{same}$ would indicate configurations with both holes on the same leg). Further peaks in amplitude appear at larger separations as the hole density decreases
[\fref{fig:results3}(i)], %
consistently favouring separations where hole exchange yields bosonic statistics either directly or with only a small perturbation, e.g.~for $6^\mrm{opp}$: \raisebox{-3.33pt}{\includegraphics[width=1.5cm]{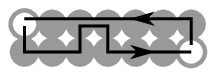}}.

In the vicinity of the critical point, the one-hop perturbations about this preferred configuration are denoted $1^\mrm{opp}$, $3^\mrm{opp}$, and $2^\mrm{same}$. %
As shown in \fref{fig:results3}(ii),
the probabilities for these configurations loosely converge in the vicinity of the critical point,
while two-hop perturbations are rarer, and
the low probability for two holes to appear 
on the same rung, $0^\mrm{opp}$, indicates that hole exchange is relatively uncommon. These results are indicative of a phase in which holes are regularly spaced and poorly delocalised, predominantly 
due to 
small fluctuations about their preferred %
separations. Finally, the $\ff{}=1$ plateau extends to lower $\mu/t$ due to %
destructive interference. %

Generalising to other anyon species $\mbb{Z}_q$, with increasing $q$ it is anticipated that the SF1/SF2 phase transition for $\ff{}<\frac{1}{2}$ will continue to migrate right and up, and occupancy in the SF1 phase will shift further towards the hard-core boson line as $q$ increases. The SF2 phase will become narrower and the SF1/SF2 and SF2/SF1 phase transitions will become coincident in the limit $q\rightarrow\infty$, recovering the hard-core boson behaviour where only SF1 behaviour is observed throughout. No lateral shift of the SF2/SF1 phase boundary was observed, and thus although the $q\rightarrow\infty$ limit of hard core bosons exhibits particle/hole symmetry, it may be that this limit is approached asymmetrically with the phase boundaries converging at some value $\mu/t>0$. %
For odd $q$ there is a novel hole-crystal phase, in which vacancies form a regularly-spaced foam %
purely as a result of the particle exchange statistics. %
The extent of this phase is expected to decrease with increasing $q$. It is characterised by hole separations consistent with bosonic exchange statistics, most notably $(\frac{q}{2}+\frac{1}{2})^\mrm{opp}$, %
implying that the filling fraction at which the SF1/HC phase transition takes place must satisfy $\ff{}\geq q/(q+1)$. This phase therefore %
vanishes in the limit $q\rightarrow\infty$. For both even and odd $\mbb{Z}_q$ the width of the $\ff{}=1$ plateau is also expected to decrease with increasing $q$, as the phase angle $R^{11}$ associated with braiding decreases and destructive interference becomes less potent.
It is conjectured that
hole-crystal phases like the one observed here could potentially also 
appear in full 2D systems, with smallest unit cells for $\mbb{Z}_3$ on the square, triangular, and hexagonal lattices appearing as follows:
\begin{center}%
\includegraphics[width=0.5\columnwidth]{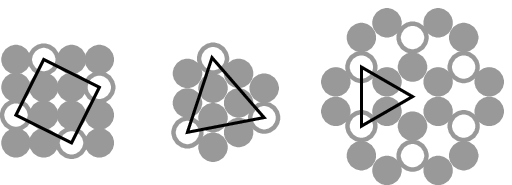}
\end{center}%

In summary, %
the phase diagram for hard-core non-interacting $\mbb{Z}_3$ anyons on the two-leg ladder is %
qualitatively distinct both from that of fermions and of hard-core bosons \cite{crepin2011}, with destructive interference giving rise to extended insulating phases at filling fractions $\ff{}=\frac{1}{2}$ and $\ff{}=1$. A novel first-order phase transition is identified at high filling fraction, into a phase characterised by a regular arrangement of strongly localised holes. %
Equivalent phases are expected to appear for all $\mbb{Z}_q|_{q\in\mbb{Z}^\mrm{odd}}$%
.
Qualitative explanations are proposed for all observed properties of the $\mbb{Z}_3$ phase diagram, with extrapolative predictions being made for %
$\mbb{Z}_q$ anyon models with ${q>3}$. These predictions should be readily amenable to testing through further numerical simulations.

\begin{acknowledgments}
The author thanks Gavin Brennen for beneficial discussions. The simulations performed in this paper made use of the \texttt{netcon} package \cite{pfeifer2014a}.
\end{acknowledgments}

\bibliography{PhaseDiag}

\end{document}